\documentstyle[12pt,a4]{article}

\begin{document}

{\Huge\bf\centerline{Two Fuzzy Logic Programming Paradoxes}}

{\Huge\bf\centerline{$\Rightarrow$}}

{\Huge\bf\centerline{Continuum Hypothesis = ``False"}}

{\Huge\bf\centerline{Axiom of Choice = ``False"}}

{\Huge\bf\centerline{$\Rightarrow$}}

\bigskip
{\Huge\bf\centerline{ZF$\not$C is Inconsistent}}

\bigskip

{\centerline {Rafee Ebrahim Kamouna}}

{\centerline {Email: rafee102000@yahoo.com}}

{\centerline {submitted to ACM Transactions on Computational Logic}}

\bigskip\bigskip
{\bf\centerline{Abstract}}
\bigskip

\noindent Two different paradoxes of the fuzzy logic programming system of [29] are presented. The first paradox is due to two distinct (contradictory) truth values for every ground atom of $FLP$, one is syntactical, the other is semantical. The second paradox concerns the cardinality of the valid $FLP$ formulas which is found to have contradictory values: both $\aleph_0$ the cardinality of the natural numbers, and $c$, the cardinality of the continuum. It follows that both {\bf CH} \& {\bf AC} are false. Hence, {\bf ZF$\not$C} is inconsistent.

\bigskip\bigskip
\noindent{\bf 1. Introduction}

\bigskip
\noindent Fuzzy logic programming and possibilistic logic programming systems in the works of Alsinet and Godo et al. [1-18], Vojtas et al. [45-49] were developed with large number of soundness and completeness results with interesting properties. Variations as the multi-adjoint logic programming was developed by Medina et al. [33-36]. The first use of truth constants in the language syntax first appeared in Pavelka's logic [41] as early as 1979. Before that, truth was expressed only in the language semantics as in Lukasiewicz and Kleene many-valued logics. Pavelka extended Lukasiewicz logic with rational truth constants. Novak [37-40], in his weighted inference systems developed a syntax of pairs: (formula, truth value). Expansions of other logics with truth constants in Esteva et al. 2000, and recently in Esteva et al. 2006 [24-27], and Savicky et al. 2006 [43]. In 2007, truth constants appeared in Esteva et al. [28]. The work of Bobillo and Straccia et al. [19,20,42,44] in fuzzy description logics employed truth constants as well. So, the idea of having a truth constant in the language syntax is well-established. This paper presents two different paradoxes as properties of the fuzzy logic programming system presented in [29].

\bigskip
\noindent In an attempt to re-organize the XX$^{th}$ significant negative results of mathematics and computation, the present (poor) author introduced the ``Syntactico-Semantical Bi-Polar Disorder Taxonomy":

\begin{enumerate}

\item Self-referential $SySBPD$:

\begin{enumerate}
\item Russell's paradox.

\item The Liar's paradox.

\end{enumerate}

\item G\"{o}del Completeness/Incompleteness $SySBPD$; note the relationship between the proof of his celebrated incompleteness theorem and the Liar's paradox.

\item Turing Decidability/Undecidability $SySBPD$.

\item Finiteness/Infiniteness $SySBPD$: results in finite model theory that succeed infinitely and fail finitely. Most importantly, G\"{o}del's completeness theorem which is:

\begin{enumerate}

\item {\em Positive}: Completeness/Incompleteness $SySBPD$.

\item {\em Negative}: Finiteness/Infiniteness $SySBPD$.

\end{enumerate}

\end{enumerate}

\noindent All these $SySBPD$'s are instances of the {\bf ``Syntactico-Semantical Precedence/Principality Bi-Polar Disorder"}.

\begin{enumerate}
\item Precedence: syntax definition precedes semantics:

[Syntax $<$ Semantics]$_{Precdence}$.

\item Principality: during computation the input takes various syntactic forms where semantics is principal over syntax in every computation step:

    [Semantics $<$ Syntax]$_{Principality}$.

\item (1) \& (2) $\Longrightarrow$ [Syntax] $<>$ [Semantics], i.e. Bi-Polar Disorder.
\end{enumerate}

\noindent The question:``Are the XX$^{th}$ the only $SySBPD$'s" led to the discovery of those two $FLP$ $SySBPD$'s. Any more $SySBPD$'s? An open question.

\bigskip

\noindent{\bf 2. The First SySBPD FLP Paradox}

\noindent An atom in fuzzy logic programming in [29] looks like:
$$p(t_1,t_2,\ldots,t_n,\mu)$$

\noindent Consider a program consisting only of a ground fact in $FLP$, e.g.:

\bigskip
$AgeAbout21(John,0.9)$

\bigskip
\noindent running this program with any ground goal results in answers either:``1" or ``0" (semantical truth value), which is in contradiction with the truth constant:``0.9" - (syntactical truth value). This is because atoms in $FLP$ are classical even when the weight is attached to them. Now this paradox is formalized rigorously in the following theorem.

\bigskip
\noindent First, the classical definition of an Herbrand interpretation and an Herbrand model are recalled. Second, it is shown that if truth constants are allowed in the language syntax in the sense of [29], then every Herbrand interpretation of any $FLP$ language is a {\em model} iff it is {\em not a model}, except for the case when $FLP$ collapses to classical logic, i.e. $\mu = ``0"$ or $\mu = ``1"$. This is the first ``Syntactico-Semantical Bi-Polar Disorder $FLP$ Paradox".

\bigskip\bigskip

\noindent {\bf Definition 1:} Let $L$ be a language over an alphabet $\Sigma$ containing at least one constant symbol. The set $U_L$ of all ground terms constructed from functions and constants in $L$ is called the Herbrand universe of $L$. The set $B_L$ of all ground atomic formulas over $L$ is called the Herbrand base of $L$.

\bigskip
\noindent {\bf Definition 2:} The {\em Herbrand interpretation} $I_L$ for a language $L$ is a structure $I_L\equiv <I_c,I_f,I_p>$ whose domain of discourse is $U_L$ where:

\begin{enumerate}

\item $\forall c\in L:c$ is a constant:
$$I_c(c) = c$$.

\item $\forall f\in L:f$ is a function symbol of arity $n$, and $t_1,t_2,\ldots,t_n$ are terms:
$$I_f(f)(t_1,t_2,\ldots,t_n) = f(I(t_1),\ldots,I(t_n))$$

\item $\forall p\in L:p$ is a predicate of arity $n$:
$$I_p(p):B_L\rightarrow\{0,1\}$$

\end{enumerate}

\bigskip
\noindent {\bf Definition 3:} The {\em Herbrand interpretation} $I_L$ for a language $L$ is a model iff $I_L:B_L\rightarrow \{1\}\wedge B_L\not\rightarrow \{0\}$

\bigskip
\noindent Let $L$ be the classical logic program consisting of the single (ground) fact:
$$p(c_1,c_2,\ldots,c_n)\leftarrow$$
\noindent and let $c_n=\mu\in C\subseteq [0,1]$ be a truth constant. If $I_L$ is an Herbrand interpretation for $L$, then $I_L$ is a {\em model} iff it is {\em not a model} (unless $\mu =``0"$ or $``1"$, i.e. $FLP$ collapses into classical logic). $I_L$ interprets the predicate symbol $p$ (classically) as a relation between the domains from which the n-tuple $(c_1,c_2,\ldots,c_n)$ is extracted. The last member of the tuple $c_n$ is a real number in a countable $C\subseteq [0,1]$. When constant symbols are interpreted in classical semantics, it banishes an argument of a predicate to be the truth constant of the same predicate. $FLP$ non-classical semantics enforces an argument of a predicate to be a truth constant of the same predicate. Semantics of formal languages are enforced in the same way as in natural languages. Since the string ``main" over the Latin alphabet is interpreted differently in English and French (the word ``main" in French means ``hand"). Obviously,

$$Oxford(main) \neq Larousse(main)$$
$$I_{L_{Classical}}\ (p) \not\equiv I_{L_{FLP}}\ (p)$$

\bigskip
\noindent Neither the English people may ask the French to follow Oxford dictionary, nor the French may ask the English to follow Larousse. Forbidding arguments of a predicate to be the truth constant of the same predicate is equally {\bf unacceptable}. Moreover, if someone attempts to attack the {\bf P} vs. {\bf NP} question by examining the properties of any language, he may do so. The entire scientific community is pre-occupied with ANY set of strings (a language) that may separate the two classes. Usually, a set of strings in {\bf NP} and not in {\bf P}, hence the question is settled. Let alone the self-referential nature of the question, i.e. {\bf P} vs. {\bf NP} is a question in {\bf NP}. So, if $X$ is the decision problem $X\equiv$ {\bf P} =? {\bf NP}, then $X\in$ {\bf NP}. But classes are (forbidden) to be elements, so such an argument is a meta-mathematical/philosophical one ($X$ is not a valid mathematical object). Just consider an analogy of the question: $x?=y, x\in N, y\in R$. Obviously, this later question is an ill-posed one. This situation encourages researchers to investigate any family of languages for possible potential important implications.

\bigskip
\noindent For the above considerations, the author is not deterred to enforce such semantics on the same syntax of classical logic, then examine the consequences. Forbidding such semantics won't help because both classes contain infinite number of languages. Any method to forbid such semantics can obviously be eliminated with a counter-part to enforce whatever semantics to examine its implications to this long outstanding question. In other words, a counter-argument against $FLP$ non-classical semantics should prove that such languages don't exist at all. The fact that it leads to paradoxical and inconsistent computations never means that these computations are wrong or meaningless. $FLP$ meta-interpreters have been implemented and work quite well meaningfully from a practical engineering point-of-view. The reason for this is that in a logic programming system, the user is interested in answer substitutions rather than logical consequences as in automatic theorem proving. Cantor's set theory has its famous paradoxes, one can never argue it is wrong, though initially it was controversial. The following theorem proves that languages written in $FLP$ can have interpretations consisting of paradoxical structures.

\bigskip
\noindent{\bf Theorem 1:} Let $L$ be the classical logic program consisting of the single (ground) fact:
$$p(c_1,c_2,\ldots,c_n)\leftarrow$$
\noindent and let $c_n=\mu\in [0,1]$ be a truth constant (with the countability restriction). If $I_L$ is an Herbrand interpretation for $L$, then $I_L$ is a {\em model} iff it is {\em not a model} (unless $\mu =``0"$ or $``1"$, i.e. $FLP$ collapses into classical logic).

\bigskip
\noindent{\bf Proof:}

\begin{enumerate}

\item $I_L\ \equiv\ <I_c,I_f,I_p>\equiv <I_c,I_p>$.

\item $\Rightarrow I_c(c_1) = c_1$.

\item $\Rightarrow I_c(c_2) = c_2$.

\item $\cdots$

\item $\cdots$

\item $\cdots$

\item $\Rightarrow I_{c_{n-1}}=c_{n-1}$.

\item $\Rightarrow I_c(\mu) = \mu\in [0,1]$.

\item $\Rightarrow I_p\in\{0,1\}$.

\item $\Rightarrow I_L\ \equiv\ <I_c,I_p>$.

\item $\Rightarrow I_L\ \equiv\ <I_c\in [0,1],I_p\in\{0,1\}>$

\item $\Rightarrow$ $\forall I_c\in ]0,1[, I_L$ is a {\em model} iff it is {\em not a model}  $\ \rule{2mm}{2mm}$

\end{enumerate}

\noindent{\bf 3. The Second SySBPD FLP Paradox}

\noindent For a program $L$ written in $FLP$, what is the cardinality of $Valid(L)$, the valid statements of $L$. Considering G\"{o}del completeness theorem for predicate calculus one has:
$$|Valid(L)|_{L:Classical}=\aleph_0$$
\noindent where $\aleph_0$ is the cardinality of the natural numbers. But after lifting the countability restriction on $\mu\in C\subseteq [0,1]$, one has:
$$|Valid(L)|_{L:Fuzzy}=c$$
\noindent where $c$ is the cardinality of the continuum. So the second $SySBPD$ $FLP$ paradox can be expressed as:
$$[|Valid(L)|=\aleph_0]_{Classical}\ \Longleftrightarrow\ [|Valid(L)|=c]_{Fuzzy}$$

\noindent The implications of those paradoxes would be considered by computer scientists/mathematicians in general and computational complexity theorists in particular.Let $P, Q, R$ and $S$ be as follows:

\begin{enumerate}

\item $P = I(p)\in\{0,1\}$.

\item $Q = I(p)\in [0,1]$.

\item $R = |Valid(L)|\ =\ \aleph_0$.

\item $S = |Valid(L)|\ =\ c$, then:

\item Paradox I: (Theorem 1)

{\centerline {$I(p)\in\{0,1\}\ \Longleftrightarrow\ I(p)\in [0,1]]\ \ \equiv\ \ [P\Longleftrightarrow Q]$}}

\item Paradox II: (Conjecture)

{\centerline {$|Valid(L)|=\aleph_0\ \Longleftrightarrow\ |Valid(L)|=c\ \ \equiv\ \ [R\Longleftrightarrow S]$}}

\end{enumerate}

\bigskip
\noindent{\bf Theorem 2:} Paradox I $\Longrightarrow$ Paradox II.

\bigskip
\noindent{\bf Proof:}

\begin{enumerate}

\item $P\Longleftrightarrow Q$.

\item $[I(p)\in\{0,1\}\Longleftrightarrow |Valid(L)| = \aleph_0] \ \ \equiv\ \ [P\Longleftrightarrow R]$.

\item $[I(p)\in [0,1]\Longleftrightarrow |Valid(L)| = c] \ \ \equiv\ \ [Q\Longleftrightarrow S]$.

\item $P\Longleftrightarrow S$, (1) \& (3).

\item $R\Longleftrightarrow S \equiv$ Pardox II, (2) \& (4), Q.E.D. \ \ \ $\rule{2mm}{2mm}$

\end{enumerate}

\noindent Let $L$ be a fuzzy program written in $FLP$ whose fuzzy
atom $p(t_1,t_2,\ldots,t_n,\mu)$ is {\em fuzzy} iff it is {\em not fuzzy}. The
notation $|Valid(L)|$ denotes the cardinality of the class of valid
formulas of $L$. Since $FLP$ is classical, we are sure that $|Valid(L)|=\aleph_0$. But we know that $FLP$ is (paradoxically)
fuzzy, then $|Valid(L)|=\aleph_0$ only if $\mu\in C\subseteq [0,1]$ where $C$ is
countable. If this condition is lifted (which is the interesting case here), i.e. $[0,1]$ is
taken to be uncountable, we would have $|Valid(L)|=c$. So, (paradoxically):
$$|Valid(L)|=\aleph_0\ \Longleftrightarrow\ |Valid(L)|=c$$

\noindent where $c$ is the cardinality of the continuum.
The following proof directly demonstrates that
the Continuum Hypothesis is ``False" and $ZFC$ is inconsistent.

\bigskip
\noindent{\bf Theorem 3:} Paradox I $\Longrightarrow$ Paradox II $\Longrightarrow$ {\bf ZFC} is inconsistent.

\begin{enumerate}

\item $\exists\ \alpha = |Valid(L)|: \aleph_0 < \alpha < c.$

\item  $[\aleph_0 < \alpha < c\Rightarrow\neg CH] \Rightarrow ZFC$ is Inconsistent.

\end{enumerate}

\bigskip
\noindent{\bf Proof:}

\noindent The proof is as follows:

\begin{enumerate}

\item $p$ is {\em fuzzy} $\ \Longleftrightarrow\ p\ $ is {\em not fuzzy}.

\item $\Rightarrow [(\alpha=\aleph_0)\ \Longleftrightarrow\ (\alpha=c)$].

\item $[(\alpha = \aleph_0)\ \Rightarrow\ (\alpha = c)]\ \Rightarrow\ \alpha\neq\aleph_0$.

\item $[(\alpha < \aleph_0)\ \wedge\ (\not\exists\ \alpha < \aleph_0)]\ \Rightarrow\ \alpha\not < \aleph_0$.

\item $[(\alpha\neq\aleph_0)\ \wedge\ (\alpha\not < \aleph_0)]\ \Rightarrow\ \alpha > \aleph_0$.

\item $[(\alpha = c)\ \Rightarrow\ (\alpha =\aleph_0)]\ \Rightarrow\ \alpha\neq c$.

\item $[(\alpha > c)\ \wedge (\not\exists\ \alpha > c)]\ \Rightarrow\ \alpha < c$.

\item $\Rightarrow \aleph_0 < \alpha < c$.

\item $\Rightarrow \neg CH$.

\item $ZFC$ is consistent $\Rightarrow CH$ is formally independent (G$\ddot{o}$del [133] \& Cohen [48]).

\item $\neg CH\Rightarrow CH$ is formally dependent.

\item $\Rightarrow ZFC$ is Inconsistent!  $\rule{2mm}{2mm}$

\end{enumerate}

\noindent Let $\Sigma$ be the alphabet of the $FLP$ system and $\Sigma^*$ be the set of all finite strings over $\Sigma$, then the set of all languages $L\subseteq \Sigma^*$ is uncountable. For each $L$, a paradoxical cardinal $\alpha$ is associated with the set of all valid formulas of $L$.

\bigskip
\noindent{\bf Theorem 4:} Axiom of Choice = ``False".

\begin{enumerate}

\item $\forall L\ \exists\alpha:\alpha$ is countable iff it is uncountable.

\item The set of all paradoxical cardinals CANNOT be well-ordered.

\item $\Longrightarrow$ Axiom of Choice = ``False".   $\rule{2mm}{2mm}$

\end{enumerate}

\break

\bigskip
\noindent The following proof emphasizes and clarifies the above results.

\bigskip
\noindent{\bf Theorem 5:} Let $p(t,\mu)$ be an atomic formula written in $FLP$, there are four cases:

\begin{enumerate}

\item t:countable, $\mu$ countable, then irrelevant to CH \& AC, i.e. only Paradox I.

\item t:countable, $\mu$ uncountable: uncountable pairs.

\begin{enumerate}

\item $\forall\ \mu\  \exists\ p.$

\item $\Longrightarrow\ |p|=c.$

\item $\forall\ \mu\ \in\ ]0,1[\ \not\exists\ Valid(p)_{classical}.$

\item $\forall\ \mu\ \in\ \{1\}\ \exists\ Valid(p)_{classical}.$

\item $\Longrightarrow\ |Valid(p)|_{classical}\neq c.$

\item $\forall\ \mu\ \in\ ]0,1]\ \exists\ Valid(p)_{fuzzy}.$

\item $\Longrightarrow\ |Valid(p)|_{fuzzy}=c.$

\item (e) \& (g) $\Longrightarrow$ the set of valid formulas is countable iff it is uncountable, Paradox II.

\end{enumerate}

\item t:uncountable, $\mu$:countable: uncountable pairs

\begin{enumerate}

\item $\forall\ \mu\  \exists\ p.$

\item $\Longrightarrow\ |p|=c.$

\item $\forall\ \mu\ \in\ \{1\}\ \exists\ Valid(p)_{classical}.$

\item $\Longrightarrow\ |Valid(p)|_{classical}\neq c.$

\item $\forall\ \mu\ \in\ ]0,1]\ \exists\ Valid(p)_{fuzzy}.$

\item $\Longrightarrow\ |Valid(p)|_{fuzzy}\neq c.$

\item (e) \& (g) $\Longrightarrow$ No Paradox II.

\end{enumerate}

\item t:uncountable, $\mu$:uncountable: uncountable pairs, i.e. same as (2) above.

\end{enumerate}

\break
\noindent{\bf References:}
\begin{enumerate}

\item T. Alsinet, L. Godo:``Adding similarity-based reasoning capabilities to a Horn fragment of possibilistic logic with fuzzy constants". Fuzzy Sets and Systems 144(1): 43-65 (2004)
2003.

\item T. Alsinet, C. Ans\'{o}tegui, R. Béjar, C. Fern\'{a}ndez, F. Manyà:``Automated monitoring of medical protocols: a secure and distributed architecture". Artificial Intelligence in Medicine 27(3): 367-392 (2003).

\item T. Alsinet, R. Béjar, A. Cabiscol, C. Fern\'{a}ndez, F. Manyà:``Minimal and Redundant SAT Encodings for the All-Interval-Series Problem. CCIA 2002: 139-144.

\item T. Alsinet, L. Godo, S. Sandri:``Two formalisms of extended possibilistic logic programming with context-dependent fuzzy unification: a comparative description". Electr. Notes Theor. Comput. Sci. 66(5): (2002).

\item T. Alsinet, L. Godo:``Towards an automated deduction system for first-order possibilistic logic programming with fuzzy constants". Int. J. Intell. Syst. 17(9): 887-924 (2002)
2001.

\item T. Alsinet, L. Godo: ``A Proof Procedure for Possibilistic Logic Programming with Fuzzy Constants". ECSQARU 2001: 760-771
2000.

\item T. Alsinet, R. Béjar, C. Fernandez, F. Manyà:``A Multi-agent system architecture for monitoring medical protocols". Agents 2000: 499-505.

\item T. Alsinet, L. Godo:``A Complete Calcultis for Possibilistic Logic Programming with Fuzzy Propositional Variables". UAI 2000: 1-10
1999.

\item T. Alsinet, L. Godo, S. Sandri:``On the Semantics and Automated Deduction for PLFC, a Logic of Possibilistic Uncertainty and Fuzziness". UAI 1999: 3-12.

\item T. Alsinet, F. Manyà, J. Planes:``Improved Exact Solvers for Weighted Max-SAT". SAT 2005: 371-377
2004.

\item T. Alsinet, F. Manyà, J. Planes:``A Max-SAT Solver with Lazy Data Structures". IBERAMIA 2004: 334-342.

\item T. Alsinet, C. I. Chesnevar, L. Godo, G. R. Simari: ``A logic programming framework for possibilistic argumentation: Formalization and logical properties". Fuzzy Sets and Systems 159(10): 1208-1228 (2008)

\item T. Alsinet, C. I. Chesnevar, L. Godo, G. R. Simari: ``A logic programming framework for possibilistic argumentation: Formalization and logical properties", Fuzzy Sets and Systems 159(10): 1208-1228 (2008).

\item T. Alsinet, L. Godo:``Adding similarity-based reasoning capabilities to a Horn fragment of possibilistic logic with fuzzy constants". Fuzzy Sets and Systems 144(1): 43-65 (2004)

\item T. Alsinet, L. Godo, S. Sandri: Two formalisms of extended possibilistic logic programming with context-dependent fuzzy unification: a comparative description. Electr. Notes Theor. Comput. Sci. 66(5): (2002)
\item T. Alsinet, L. Godo: Towards an automated deduction system for first-order possibilistic logic programming with fuzzy constants. Int. J. Intell. Syst. 17(9): 887-924 (2002)

\item T. Alsinet, L. Godo: A Proof Procedure for Possibilistic Logic Programming with Fuzzy Constants. ECSQARU 2001: 760-771

\item T. Alsinet, L. Godo: A Complete Calcultis for Possibilistic Logic Programming with Fuzzy Propositional Variables. UAI 2000: 1-10

\item F. Bobillo and U. Straccia:``On Qualified Cardinality Restrictions in Fuzzy Description Logics under Lukasiewicz semantics". In Proceedings of the 12th International Conference on Information Processing and Management of Uncertainty in Knowledge-Based Systems, (IPMU-08), 2008.

\item F. Bobillo and U. Straccia:``fuzzyDL:An Expressive Fuzzy Description Logic Reasoner". In Proceedings of the 2008 International Conference on Fuzzy Systems (FUZZ-08).

\item C. I. Chesnevar, Guillermo Ricardo Simari, Lluis Godo, Teresa Alsinet: ``Argument-Based Expansion Operators in Possibilistic Defeasible Logic Programming: Characterization and Logical Properties". ECSQARU 2005: 353-365.

\item C. I. Chesnevar, Guillermo Ricardo Simari, Lluis Godo, Teresa Alsinet: Expansion Operators for Modelling Agent Reasoning in Possibilistic Defeasible Logic Programming. EUMAS 2005: 474-475

\item C. I. Chesnevar, Guillermo Ricardo Simari, Teresa Alsinet, Lluis Godo:``A Logic Programming Framework for Possibilistic Argumentation with Vague Knowledge". UAI 2004: 76-84.

\item F. Esteva, L. Godo:``Putting together Lukasiewicz and product logic", Mathware and Soft Computing 6:219:234, 1999.

\item F. Esteva, L. Godo, P. Hajek and M. Navara:``Residuated Fuzzy Logics with an Involutive Negation", Archive for Math. Log., 39: 103-124.

\item F. Esteva, L. Godo:``Monoidal t-norm Based Logic", Fuzzy Sets and Systems, 124:271-288, 2001.

\item F. Esteva, L. Godo and C. Noguera:``On Rational Weak Nilpotent Minimum Logics", J. Multiple-Valued Logic and Soft Computing, 2006.

\item F. Esteva, J. Gispert, L. Godo, C. Noguera:``Adding Truth-Constants to Logics of Continuous t-norms: Axiomatization and Completeness Results", Fuzzy Sets and Systems, 158:597-618, 2007.

\item R. E. Kamouna: ``Fuzzy Logic Programming", Fuzzy Sets and Systems, 1998.

\item S. Krajci, R. Lencses, P. Vojt\'{a}s:``A comparison  of fuzzy and annotated logic programming". Fuzzy Sets and Systems, 144 (2004) 173–192

\item T. Lukasiewicz and U. Straccia:``Managing Uncertainty and Vagueness", in Description Logics for the Semantic Web In Journal of Web Semantics.

\item S. Krajci, R. Lencses, P. Vojt\'{a}s:``A comparison  of fuzzy and annotated logic programming". Fuzzy Sets and Systems, 144 (2004) 173–192

\item J. Medina, M. Ojeda, P. Vojt\'{a}s:``Multi-adjoint logic programming with continuous semantics". In Proc. LPNMR'01. Th. Eiter et al eds. Lecture Notes in Artificial Intelligence 2173, Springer Verlag 2001, 351-364

\item J. Medina, M. Ojeda, P. Vojt\'{a}s:``A procedural semantics for multi-adjoint logic programming. In Proc. EPIA'01, P. Brazdil and A. Jorge eds. Lecture Notes in Artificial Intelligence 2258, Springer Verlag 2001, 290-297

\item J. Medina, M. Ojeda, P. Voj\'{a}s:``A completeness theorem for multi-adjoint logic programming". In Proc. 10th IEEE  Internat. Conf. Fuzzy Systems, IEEE 2001, 1031-1034,

\item J. Medina, M. Ojeda-Aciego, A. Valverde, P. Vojt\'{a}s:``Towards Biresiduated Multi-adjoint Logic Programming". R. Conejo et al Eds. Revised Selected Papers of CAEPIA 2003. Lecture Notes in Computer Science 3040 Springer 2004, 608-617,

\item V. Novak, I. Perfilieva and J. Mockor: ``Mathematical principles of fuzzy logic", Kluwer, Boston/Dordrecht, 1999.

\item V. Novak:``Weighted inference systems", in J. C. Bezdek, D. Dubois and H. Prade (eds.): Fuzzy Sets in Approximate Reasoning and Information Systems. Handbooks of Fuzzy Sets Series, Vol. 3. Kluwer, Boston, 191-241, 1999.

\item V. Novak and I. Perfilieva (eds.):``Discovering the World with Fuzzy Logic; Studies in fuzziness and soft computing", Heidelberg, New York: Physica-Verlag, Vol. 57, 302-304, 2000.

\item V. Nov\'{a}k, S. Gottwald, P. H\'{a}jek: Selected papers from the International Conference "The Logic of Soft Computing IV" and Fourth workshop of the ERCIM working group on soft computing. Fuzzy Sets and Systems 158(6): 595-596 (2007)

\item J. Pavelka:``On Fuzzy Logic I-III. Zeit", Math Logik Grund. Math. 25, 45-52, 119-134, 447-464, 1979.

\item A. Ragone, U. Straccia, T. Di Noia, E. Di Sciascio and F. M. Donini:``Fuzzy Description Logics for Bilateral Matchmaking in e-Marketplaces". In Proceedings of the 16th Italian Symposium on Advanced Database Systems (SEBD-08), 2008.

\item P. Savicky, R. Cignoli, F. Esteva, L. Godo, C. Noguera:``On Product Logic with Truth-constants, Journal of Logic and Computation, Volume 16, Number 2, pp. 205-225(21), Oxford University, 2006.

\item U. Straccia:``Fuzzy Description Logic Programs", in Uncertainty and Intelligent Information Systems, B. Bouchon-Meunier, R.R. Yager, C. Marsala, and M. Rifqi eds. , 2008.

\item P. Vojt\'{a}s:``Fuzzy logic programming". Fuzzy Sets and Systems. 124,3 (2001) 361-370

\item P. Vojt\'{a}s, T. Alsinet, Ll. Godo:``Different models of fuzzy logic programming with fuzzy unification (towards a revision of fuzzy databases)". In Proc. IFSA'01 Vancouver, IEEE, 2001, 1541-1546,

\item P. Vojt\'{a}s:``Tunable fuzzy logic programming for abduction under uncertainty". In Proc. Workshop Many Valued Logic for Computer Science Applications. European Conference on Artificial Intelligence 98, University of Brighton, 1998, 7 pages

\item P. Vojt\'{a}s. L. Paulak:``Soundness and completeness of non-classical extended SLD-resolution", in Proc. ELP'96 Extended logic programming, Leipzig, ed. R. Dyckhoff et al., Lecture Notes in Comp. Sci. 1050 Springer Verlag, 1996, 289-301.

\item P. Vojt\'{a}s, M. Vomlelov\'{a}:``Transformation of deductive and inductive tasks between models of logic programming with imperfect information", In Proc. IPMU 2004, B. Bouchon-Meunier et al. eds. Editrice Universita La Sapienza, Roma, 2004, 839-846

\end{enumerate}

\end{document}